\documentclass[showpacs,preprintnumbers,amsmath,amssymb,floatfix]{revtex4}

\usepackage{graphicx}
\usepackage{dcolumn}
\usepackage{bm}

\begin{document}

\def\vv{{\hbox{\boldmath $v$}}}
\def\bbeta{{\hbox{\boldmath $\beta$}}}
\def\bkappa{{\hbox{\boldmath $\kappa$}}}


\title{HBT interferometry with quantum transport of the interfering pair
}

\author{Li-Li Yu$^1$}
\author{Wei-Ning Zhang$^{1,2}$}
\author{Cheuk-Yin Wong$^{2,3}$}


\affiliation{
$^1$Department of Physics, Harbin Institute of Technology,
Harbin, Heilongjiang 150006, P. R. China\\
$^2$School of Physics and Optoelectronic Technology, Dalian University of 
Technology, Dalian, Liaoning 116024, P. R. China\\
$^3$Physics Division, Oak Ridge National Laboratory, Oak Ridge, TN
37831, U.S.A.
}

\date{\today}
\begin{abstract}
In the late stage of the evolution of a pion system in high-energy
heavy-ion collisions when pions undergo multiple scatterings, the
quantum transport of the interfering pair of identical pions plays an
important role in determining the characteristics of the
Hanbury-Brown-Twiss (HBT) interference.  We study the quantum
transport of the interfering pair using the path-integral method, in
which the evolution of the bulk matter is described by relativistic
hydrodynamics while the paths of the two interfering pions by test
particles following the fluid positions and velocity fields.  We
investigate in addition the effects of secondary pion sources from
particle decays, for nuclear collisions at AGS and RHIC energies.  We
find that quantum transport of the interfering pair leads to HBT radii
close to those for the chemical freeze-out configuration. Particle
decays however lead to HBT radii greater than those for the chemical
freeze-out configuration.  As a consequence, the combined effects give
rise to HBT radii between those extracted from the chemical freeze-out
configuration and the thermal freeze-out configuration.  Proper
quantum treatments of the interfering pairs in HBT calculations at the
pion multiple scattering stage are important for our understanding of
the characteristics of HBT interferometry in heavy-ion collisions.

\end{abstract}


\maketitle

\section{Introduction}

Two-pion Hanbury-Brown-Twiss (HBT) interferometry has been widely used
in high energy heavy ion collisions to study the space-time geometry
of particle-emitting sources \cite{Won94,Wie99,Wei00,Lis05}.  As is
well known, the HBT interference occurs for a chaotic source and is
absent for a coherence source \cite{Gla63}.  It is therefore important
to identify the proper ``chaotic'' source to compare theoretical
results with experimental measurements.

Usual theoretical treatments of the HBT problem argue that the
collisions in the thermalisation of pions at the late stage of the
evolution behaves as collisions of classical particles in a chaotic
system, and these chaotic collisions will lead to a chaotic thermal
freeze-out configuration.  According to this argument, the thermal
freeze-out configuration should therefore be the proper ``chaotic''
source in HBT measurements and the HBT radii are expected to increase
substantially with increasing collision energies.  However,
experimental data indicate that there are only relatively small changes
of HBT radii when the energy increases from AGS, SPS, to RHIC energies
\cite{Lis00,Agg00,Bea00,Ahl02,Sta01,Phe02,Phe04,Sta05}.

As the proper identification of the HBT ``chaotic'' pion source is
crucial to the understanding of the phenomenon, it is necessary to
examine the nature of the pion source during the late stage of its
evolution.  One envisages that as the pion system cools down, the
kinetic energies of the pions may initially be quite high so that
chemical reactions among pions can take place, leading to the
conversion of pions into kaons (and vice versa).  These chemical
reactions will lower their intensities significantly when the
temperature decreases below a limit. The system will undergo chemical
freeze-out at which the yield ratios of dominant particles are nearly
fixed.  Before chemical freeze-out, the identities of the particles
are still in a state of flux.  Therefore, it is reasonable to take the
chemical freeze-out configuration to be the initial ``chaotic'' pion
source for the purpose of studying the HBT interference.

The ``chaotic'' source is however not a static source from which the
pair of interfering identical pions are emitted and detected.  The
chaotic source continues to evolve and the pair of pions must follow
the evolution from the chemical freeze-out configuration to the
state of thermal freeze-out.  Only at the state of thermal freeze-out
can the pair of pions be free of interactions and can be considered
``emitted'' to the detector.  How the pair of pions propagate from the
initial chemical freeze-out point to the thermal freeze-out point will
affect the characteristics of HBT interference and will be the subject
of the present investigation.

At temperatures below the chemical freeze-out temperature, pions
undergo multiple scattering with the medium particles.  The scattering
can be elastic or inelastic. Thermalisation takes place until the
temperature reach the thermal freeze-out limit, at which the shapes of
momentum distributions of final particles are determined
\cite{Hir02,Shu99,Hei99}.  Thus, the last stage of the evolution of
the pion system makes a transition from a chemical-reaction dominating
epoch before chemical freeze-out to a multiple scattering dominating
epoch between chemical freeze-out and thermal freeze-out.

In the usual treatment of the HBT problem, the last stage of multiple
scattering is envisaged as random cascades of classical binary
collisions and the spatial configuration of the system after these
random binary collisions will continue to be chaotic in the
newly-evolved and expanded configuration.  Such a viewpoint may
indeed be valid for some other considerations of the state of the
system, it is nonetheless inadequate for the HBT interference for a
pair of identical particles.  It is important to note that the HBT
interference phenomenon is quantum mechanical in nature, arising from
the interference of pairs of identical pions traveling through
different history paths to reach two detectors.  The propagation of a
pair of pions which undergo multiple scattering with the medium at the
late stage of the evolution should be treated quantum mechanically
\cite{Won03,Won04,Won05,Zha04c,Kap04,Mil05}.

Upon carrying out such an investigation for a pair of test particles
in the path-integral method, it was previously realized that the
collisions of the pion with medium particles at the late stage of the
evolution lead to the accumulation of the phases of the pion wave
functions.  The real parts of these phase shifts of the pair tend to
cancel while the imaginary part leads to an absorption in the two-pion
interferometry \cite{Won03,Won04,Won05,Zha04c}.  As a consequence, the
effective source size is not necessarily that of the newly-evolved
spatial configuration as in thermal freeze-out, but may be
significantly smaller.  We would like to demonstrate such an effect
explicitly here by assuming a chaotic source at chemical freeze-out
and by following the quantum transport of the interfering pair, using
the path integral method for which the pions need not follow
straight-line trajectories.  For such a study we use relativistic
hydrodynamics to describe the evolution of the bulk matter while the
paths of the two interfering pions are described by test particles
following the hydrodynamical fluid positions and velocity fields.
Alternative descriptions of particle emission using different view
points have also been presented previously \cite{Kap04,Mil05,Soc04}.

In following the evolution of pairs of identical pions, there are
additional complications.  Pions are produced not only directly at
chemical freeze-out but also indirectly by particle decays as
secondary pions during the period from chemical freeze-out to thermal
freeze-out.  It is necessary to include these secondary pions in our
HBT calculations.  The HBT radius for the evolving source will reflect
not only the geometry of the source at a configuration close to the
chemical freeze-out configuration but the effects of these secondary
sources.  Because the secondary sources are different for AGS
collisions and RHIC collisions, we shall study the effects of particle
decays for collisions at both AGS and RHIC energies.  As the present
investigation is for demonstration purposes, we shall limit our
attention on spherical geometry.  To assist future investigations
utilizing quantum transport of the interfering pair, we present a
detail description of the procedures how this quantum transport is
carried out in the path integral method.

The paper is organized as follows. In Sec. II, we give a brief review
of equations of relativistic hydrodynamics and discuss the equations
of state for sources produced in collisions at AGS and RHIC energies.
We also discuss the particle production conditions in this section.
In Sec. III, we review the two-pion HBT correlation function in
quantum probability amplitudes in a path-integral formalism.  We
examine the HBT radii for the evolving source, and investigate the
effects of particle decay and multiple scattering on HBT radii both
for AGS and RHIC energies.  Finally, a summary and discussion are
presented in Sec. IV.

\section{Evolution of the Particle Source}

As a completely quantum mechanical description of the evolution of the
full system is not within reach, we shall study the evolution of the
bulk strongly-interacting matter by relativistic hydrodynamics, which
has been quite successful in high-energy heavy-ion collisions.  We
shall then examine the interference of the pair of pions by following
their trajectories as test particles in the fluid after the chemical
freeze-out.

\subsection{Equations of relativistic hydrodynamics}

In relativistic hydrodynamics, the dynamics of the
strongly-interacting fluid is defined by local conservations of
energy-momentum and net charges \cite{Ris96,Ris98,Bol03}.  In our
model calculations we use conservations of energy-momentum, net baryon
number, and entropy of the system.  The continuity equations of these
conservations are \cite{Ris96,Ris98,Bol03}
\begin{eqnarray}
\label{Tuv}
\partial_{\mu} T^{\mu\nu}(x)=0 \,,
\end{eqnarray}
\begin{eqnarray}
\label{bnc}
\partial_{\mu}j^{\mu}_{\rm b} (x)=0 \, ,
\end{eqnarray} 
\begin{eqnarray}
\label{sc}
\partial_{\mu}j^{\mu}_s (x)=0 \, ,
\end{eqnarray} 
where $x$ is the space-time coordinate of a thermalize fluid element
in the source center-of-mass frame, $T^{\mu\nu}(x)$ is the energy
momentum tensor of the element, $j^{\mu}_{\rm b} (x)=n_{\rm
b}(x)u^{\mu}$ and $j^{\mu}_s (x)=s(x)u^{\mu}$ are the
four-current-density of baryon and entropy ($n_{\rm b}$ and $s$ are
baryon density and entropy density), and $u^{\mu}=\gamma (1,\vv)$
is the 4-velocity of the fluid element.  The energy momentum tensor
can be expressed as \cite{Lan59,Ris96,Ris98,Bol03}
\begin{equation}
\label{tensor} T^{\mu \nu} (x) = \big [ \epsilon(x) + p(x) \big ]
u^{\mu}(x) u^{\nu}(x) - p(x) g^{\mu \nu} \, ,
\end{equation}
where $p$ and $\epsilon$ are the pressure and energy density of the fluid 
element, and $g^{\mu \nu}$ is the metric tensor.  

From the local conservations Eqs.\ (\ref{Tuv})--(\ref{sc}), one can
get the equations of motion for spherical geometry as
\cite{Ris96,Ris98}
\begin{eqnarray}
\label{eqe}
\partial_t E + \partial_r [(E+p)v] = - \frac{2 v}{r} (E+p) \,, 
\end{eqnarray}
\begin{eqnarray}
\label{eqm}
\partial_t M + \partial_r (Mv+p) = - \frac{2 v}{r} M \,,
\end{eqnarray}
\begin{eqnarray}
\label{eqnb}
\partial_t N_{\rm b} + \partial_r (N{\rm _b} v) = -\frac{2 v}{r} N_{\rm b}\,,
\end{eqnarray}
\begin{eqnarray}
\label{eqs}
\partial_t N_s + \partial_r (N_s v) = -\frac{2 v}{r} N_s\,,
\end{eqnarray}
where $E \equiv T^{00}$, $M \equiv T^{0r}$, $N_{\rm b}=n_{\rm b}\gamma$, 
$N_s=s\gamma$. 
Using the HALE scheme \cite{Sch93,Ris95} and Sod's operator splitting 
method \cite{Sod77}, one can obtain the solutions of Eqs. 
(\ref{eqe})--(\ref{eqs}) \cite{Ris96,Ris98,Zha04,Zha04c}, after knowing the 
initial conditions and equations of state.  In our calculations, we assume 
that the initial systems are distributed uniformly within a sphere with 
radius $r_0$ and have zero initial velocity.  The grid spacing and time 
step in the calculations are taken as $\Delta x=0.04r_0$ and $\Delta 
t=0.99 \Delta x$. 

\subsection{Equation of state}

In the equations of motion (\ref{eqe})--(\ref{eqs}), there are
$\epsilon$, $p$, $v$, $n_{\rm b}$, and $s$ five unknown functions.  In
order to obtain the solution of the equations of motion, we need an
equation of state (EOS), $p(\epsilon,n_{\rm b},s)$, which gives a
relation for $p$, $\epsilon$, $n_{\rm b}$, and $s$ \cite{Ris98}.

\subsubsection{EOS for system at AGS energies}

At AGS energies, we use a mixed perfect gas of hadrons to describe the
particle-emitting source.  The number density, energy density,
pressure, and entropy density of the particle species $i$ for a
thermalize fluid element can be expressed in the fluid element local
frame in terms of temperature $T(x)$ and chemical potential $\mu_i(x)$
as
\begin{equation}
\label{ni} 
n_i={{4\pi g_i}\over{(2\pi)^3}}\int_{m_i}^\infty f_i 
E \sqrt{E^2 \! - \! m_i^2}\,dE\,,
\end{equation}
\begin{equation}
\label{ei} 
\epsilon_i={{4\pi g_i}\over{(2\pi)^3}}\int_{m_i}^\infty f_i 
E^2 \sqrt{E^2 \! - \! m_i^2} \, dE\,,
\end{equation}
\begin{equation}
\label{pi} 
p_i={1\over3}{{4\pi g_i}\over{(2\pi)^3}}\int_{m_i}^\infty f_i
(E^2 \! - \! m_i^2)^{3/2} \, dE\, ,
\end{equation}
\begin{eqnarray}
\label{si}
s_i&=&{{4\pi g_i}\over{(2\pi)^3}}\int_{m_i}^\infty [-f_i \ln f_i  
\mp (1 \mp f_i) \nonumber \\
& & \times \ln (1 \mp f_i)] E \sqrt{E^2 \! - \! m_i^2} \, dE\, ,
\end{eqnarray}
where
\begin{equation}
\label{fi} 
f_i={1\over{\exp[(E-\mu_i)/T]\pm1}}\, ,
\end{equation}
$g_i$ and $m_i$ are the internal freedom and mass of particle species
$i$, and the sign ($+$) or ($-$) is for fermions or bosons.  The fluid 
energy density $\epsilon$, pressure $p$, and entropy density $s$ are the 
sum of $\epsilon_i$, $p_i$, and $s_i$ for all particle species considered 
in the mixed perfect gas, respectively.  The baryon density $n_{\rm b}$ 
is the sum of $n_i$ for all baryon species in the gas.

For a given set of variable $(T,\mu_1,\mu_2,\ldots)$, one can obtain the 
equation of state $p(\epsilon, n_{\rm b}, s)$ from Eqs. (\ref{ni})--(\ref{si}) 
numerically.  Additionally, using HALE scheme to solve equations of fluid 
dynamics we need the sound velocity of the fluid, $c_{\rm s}^2=\partial p/
\partial\epsilon$, which can be expressed for the mixed perfect gas as 
\begin{eqnarray}
\label{cs2}
c_{\rm s}^2 &=& \sum_i \!\int_{m_i}^\infty f_i (1\! \mp\! f_i)
(E^2\!-\!m_i^2)^{3/2}(E \! - \! \mu_i)dE \nonumber \\ 
&\div & \! ~ \!\! \sum_i \!\int_{m_i}^\infty{{f_i (1 \! \mp \! f_i)}\over 3} 
\sqrt{E^2 \! - \! m_i^2} \,(E \! - \! \mu_i) E^2 dE \,.
\end{eqnarray}

\subsubsection{EOS for system at RHIC energy} 

At RHIC energy, the system undergoes a transition from QGP phase to
hadronic phase.  The initial baryon density in the center rapidity
region of the collisions is approximately zero because the total
number of pions is much greater than the total baryon numbers and
there is some tendency of baryon transparency.  At zero net baryon
density, QCD lattice results suggest the entropy density of the system
as a function of temperature as \cite{Bla87,Lae96,Ris96,Ris98}
\begin{equation}
\label{eos} 
{s(T) \over s_{\rm c}} = \Bigg({T\over {T_{\rm c}}}\Bigg)^{\!\!3} 
\Bigg[1+{{\Theta\!-\!1} \over {\Theta\!+\!1}}\tanh\!\Bigg({{T\!-\!T_{\rm c}}
\over{\Delta T}}\Bigg)\Bigg]\,,
\end{equation}
where $s_{\rm c}$ is the entropy density at the transition temperature 
$T_{\rm c}$, $\Delta T$ ($0<\Delta T<0.1T_{\rm c}$) is the width of the 
transition, $\Theta$ is a parameter associated with the ratio of the degrees 
of freedom of the QGP phase to the hadronic phase.  For $\Delta T=0$, the 
equation of state (\ref{eos}) reduces to the MIT bag equation of state with 
bag constant $B=(\Theta-1)T_{\rm c} s_{\rm c}/2(\Theta+1)$ 
\cite{Ris96,Ris98,Cho74}.  

For zero net baryon density we only need to solve Eqs. (\ref{eqe}), 
(\ref{eqm}), and (\ref{eqs}).  The thermodynamical relations among 
$p$, $\epsilon$, and $s$ in this case are  
\begin{equation}
\label{stp} 
-s\, dT + dp=0\,,
\end{equation}
\begin{equation}
\label{esp} 
\epsilon = Ts-p \,.
\end{equation}
From these thermodynamical relations and Eq. (\ref{eos}), one can obtain 
the equation of state $p(\epsilon,s)$ and sound velocity $c_{\rm s}$.  At 
the temperature below $(1-\Delta T)T_{\rm c}$, the source is in hadronic 
phase and the equation of state is taken as the mixed perfect gas of 
hadrons with zero baryon density.  

\subsection{Particle sources}

How the particle number and its chemical composition may change as the
fluid evolves can be described by three different possibilities, some
more realistic than others.  They are all included here to show
various limiting cases.  As the equation of state of the fluid depends
on particle number and the chemical composition, they also lead to
different hydrodynamical evolutions.

\subsubsection{Partial chemical equilibrium production}

As the temperature of the fluid decreases, the kinetic energy for the
collision between particles decreases.  As a consequence, dominant
chemical reactions of converting pions to kaons will lower their
intensities and will eventually stop when the temperature decreases
below a certain limit. For the problem we are considering, the
limiting temperature occurs when the average collision energy at the
chemical freeze-out temperature is equal to the threshold for the
reaction to change two pions to two kaons.  The state of the system
when this limit of temperature is reached can be described as the
state of chemical freeze-out, when the equilibrium yield ratios of the
dominating components (pions and kaons) are frozen.

From the passage of chemical freeze-out to thermal freeze-out, the
fluid continues to evolve.  Its chemical composition can continue to
change as particles can decay into other particles with various
life-times.  Conversely, the decay product particles can recombine to
form parent particles.  Some processes with larger cross sections can
also equilibrate even below $T^{\rm ch}$.  The chemical composition
during this stage is however not in a state of complete chemical
equilibrium.  We can call it the state of partial chemical equilibrium
(PCE).  This is the most realistic case when we include the decay and
recombination of particles \cite{Hir02,Beb92}.  We follow Hirano and
Tsuda \cite{Hir02} to assume that below $T^{\rm ch}$ the chemical
potential of particle species $i$, $\mu_i(T)$, obeys
\begin{eqnarray}
\label{nsb}
\frac{\bar{n}_{i}(T,\mu_{i})}{s(T,\{\mu_{i}\})}=\frac{\bar{n}_{i}(T^{\rm ch},
\mu_{i}^{\rm ch})}{s(T^{\rm ch},\{\mu_{i}^{\rm ch}\})}\,,
\end{eqnarray}
where ${\bar{n}_{i}}=n_i +\sum_{j\ne i}{\tilde d}_{j\rightarrow
i}\,n_j$, and ${\tilde d}_{j\rightarrow i}$ is the fraction of
exited-state particle species $j$ that decays to the stable particle
species $i$ \cite{Hir02,Beb92}.

\subsubsection{Chemical freeze-out production}

For our comparison, it is of interest to examine the less realistic
cases when the decay of particles is not considered and other chemical
reactions do not take place after chemical freeze-out (CFO).  In this case,
the chemical composition of the particles are frozen as at the state
of chemical freeze-out and all final particles are produced thermally
from the source configuration at the chemical freeze-out temperature
$T^{\rm ch}<T_c$.  Below $T^{\rm ch}$ all numbers of hadrons are fixed
and that the particle number densities obey $\partial_{\mu}(n_iu^{\mu})=0$.  
Eq. (\ref{nsb}) is changed to \cite{Hir02}
\begin{eqnarray}
\label{ns}
\frac{n_{i}(T,\mu_{i})}{s(T,\{\mu_{i}\})}=\frac{n_{i}(T^{\rm ch},
\mu_{i}^{\rm ch})} {s(T^{\rm ch},\{\mu_{i}^{\rm ch}\})}\,.
\end{eqnarray}
We can obtain the chemical potential $\mu_i (T)$ as a function of temperature 
below $T^{\rm ch}$ from Eq. (\ref{ns}), which ensures keeping the number of 
the hadron $i$ fixed.  

\begin{figure}
\includegraphics[angle=0,scale=0.45]{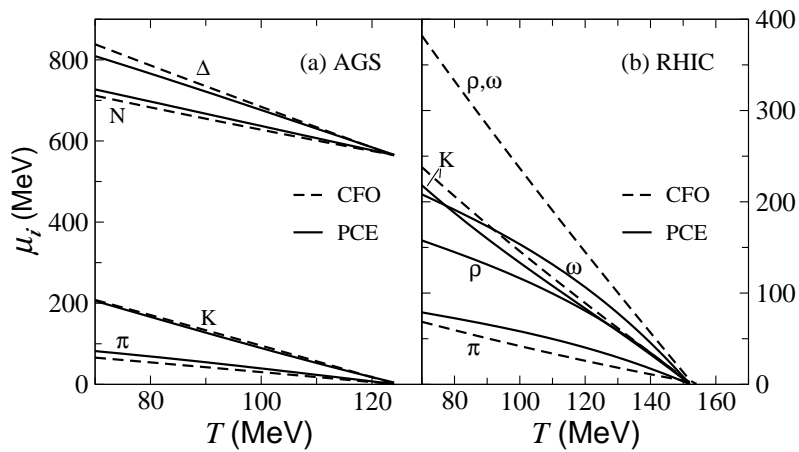}
\caption{\label{fig:fig1}Chemical potentials of hadrons for CFO production 
(dotted lines) and PCE production (solid lines) for AGS and RHIC energies.}
\end{figure}

Figs. 1(a) and (b) give the chemical potential $\mu_i(T)$ for the
sources produced at AGS and RHIC energies, and CFO and PCE modes of
production.  For the CFO production, the chemical potentials of all
particles are calculated from Eq. (\ref{ns}).  For the PCE production,
the chemical potentials of the stable particles are calculated from
Eq. (\ref{nsb}), and the chemical potentials of the excited-state
particles are obtained from the relations of chemical equilibrium,
$\mu_{\Delta}=\mu_N+\mu_{\pi}$, $\mu_{\rho}=2 \mu_{\pi}$, and
$\mu_{\omega}=3\times0.88\mu_{\pi}$ \cite{Hir02,Beb92}.

In our model calculations for the source produced in collisions at
AGS energies, the initial temperature and baryon chemical potential are
taken as 155 MeV and 540 MeV, which correspond to an initial energy
density of $\epsilon_0 =0.56$ GeV/fm$^3$.  For simplicity, we only
consider $N$, $\pi$, and $K$ as stable particles, and $\Delta(1232)$
as exited-state particle for the source with finite baryon density.
The chemical freeze-out temperature is taken as $0.8T_0=124$ MeV, and
the corresponding chemical potential is 565 MeV.  They are consistent
with the freeze-out values obtained from hadronic abundances at
CERN/SPS, NBL/AGS, and GSI/SIS \cite{Cle98}.  For the source with zero
baryon density produced in the collisions at RHIC energy, we take the
initial temperature as 250 MeV, which corresponds to an initial energy
density of $\epsilon_0=4.64$ GeV/fm$^3$.  The phase transition
temperature $T_{\rm c}$ and the width of the transition $\Delta T$ are
taken as 170 MeV and 0.05$T_{\rm c}$.  For the chemical freeze-out
temperature \cite{Hir02}, we take the chemical freeze-out temperature
as $0.9T_{\rm c} =153$ MeV.  We consider $\pi$, $K$, and $\eta$ as
stable particles and take $\rho$ and $\omega$ as excited-state
particles after the chemical freeze-out.  The value of $\Theta$ in the
equation of state (\ref{eos}) is taken as 5, by consisting the entropy
density calculated from Eq. (\ref{eos}) with the result of the entropy
density calculated for the mixed perfect gas of hadrons at the
chemical freeze-out temperature.

\subsubsection{Chemical equilibrium up to thermal freeze-out}

Finally, we can also examine another less realistic case when chemical
equilibrium can be maintained through-out the passage up to thermal
freeze-out (TFO), with the emission of particles.  Implicit in this
scenario is the assumption that all through the hydrodynamical
evolution, the kinetic energy for the collision between particles is
always large enough to allow chemical reactions to take place to
change the chemical composition.  Although this may be a good picture
for classical fluids when the temperature is comparable to the energy
needed for chemical reactions, this may not be as realistic for our
case when the kinetic energy in these temperatures is lower than
threshold energy for the reactions of two pions converting into two
kaons.

For the TFO particle production, the state of chemical freeze-out
coincides with the state of thermal freeze-out at $T^{\rm th}$.  For
AGS energies, the chemical potentials of $\pi$ and $K$ are zero and the
chemical potentials of $N$ and $\Delta$ are the same as the initial
baryon chemical potential.  For RHIC energy, all final particles have
zero chemical potential.

\section{Two-pion interferometry analysis}

\subsection{Path integral formalism of the HBT correlation function}

As the identities of particles undergo changes during chemical
reactions, the pion source at the moment of chemical freeze-out can be
characterized as chaotic in nature, for the purpose of investigating
HBT interferometry.  The system subsequently expands, its temperature
decreases, and its apparent size increases.  During the evolution from
chemical freeze-out to thermal freeze-out, the dynamics is dominated
by multiple scattering of the pions.

It is in the nature of the HBT interference that under the action of
multiple scattering after chemical freeze-out, the quantum transport
of the interfering pair of pions determines their eventual pattern of
interferometry when they reach the detector.  As a consequence, this
quantum transport has an effect on the effective source size observed
in HBT measurements.  We suggest previously that the multiple
scattering in the transport leads to an apparent source that is close
to the original chaotic source before the pair quantum transport,
supplemented by absorptions along the particle paths
\cite{Won03,Won04,Won05,Zha04c}.  In order to exhibit the importance
of this effect, we review the path integral treatment of HBT
correlation function under quantum transport as discussed earlier in
\cite{Won05}.

For definiteness, we consider the HBT interference of a pair of
$\pi^+$ particles and study first the motion of one of the two
identical pions.  As the pion propagates from the state of chemical
freeze-out to thermal freeze-out, the scattering can be described by
short-ranged scalar and vector interactions, $v_{\rm
col}^{(s,v)}(q-q_i)$, where $q$ is the coordinate of the propagating
pion and $q_i$ is the coordinate of a medium particle.  The
propagating pion is also subject to a collective flow which can be
described by a long-range density-dependent mean-field scalar and
vector interactions, $V_{\rm mf}^{(s,v)}(q)$, as in similar cases in
the dynamics of the nuclear fluid \cite{Won78}.  The Lagrangian for
the propagating pion is given by
\begin{eqnarray}
L(q,\dot q)=L_{\rm mf}(q,\dot q)+L_{\rm col}(q,\dot q),
\end{eqnarray}
\vspace*{0.3cm}where
\vspace*{-1.2cm}
\begin{eqnarray}
L_{\rm mf} (q,\dot q) = - [ m_\pi + V_{\rm mf}^{(s)}(q) ] 
                       \sqrt{ 1-{\dot {\bf q}}^2 }
+ {\dot {\bf q}} \cdot {\bf V}_{\rm mf}^{(v)}(q) - V_{\rm mf}^{0(v)}(q),
\end{eqnarray}
\vspace*{-0.6cm}
\begin{eqnarray}
L_{\rm col}(q,\dot q)=- V_{\rm col}^{(s)}(q) 
                        \sqrt{1-{\dot {\bf q}}^2}
+ {\dot {\bf q}} \cdot {\bf V}_{\rm col}^{(v)}(q) - V_{\rm col}^{0(v)}(q),
\end{eqnarray}
\vspace*{-0.8cm}
\begin{eqnarray}
\label{col}
V_{\rm col}^{(s,v)}(q)
=\sum_{i}
{v}_{\rm col}^{(s,v)}(q-q_i).
\end{eqnarray}
From the Lagrangian of the propagating pion, one obtains the pion
three-momentum,
\begin{eqnarray}
{\bf p}=\partial L/\partial {\dot {\bf q}}=
\gamma [ m_\pi+V^{(s)}(q) ] \dot {\bf q} +{\bf V}^{(v)}(q), 
\end{eqnarray}
the pion Hamiltonian, 
\vspace*{-0.3cm}
\begin{eqnarray}
H=p^0= {\bf p}\cdot {\dot {\bf q}}-L
     =\gamma [ m_\pi + V^{(s)}(q) ] + V^{0(v)}(q),
\end{eqnarray} 
and the pion mass-shell condition,
\begin{eqnarray}
[p^0-V^{0(v)}(q)]^2-[{\bf p}^2-{\bf V}^{(v)}(q)]^2 - [m_\pi+V^{(s)}(q)]^2=0,
\end{eqnarray}
where $\gamma=1/\sqrt{1-{\dot {\bf q}}^2}$ and $V^{(s,v)}(q)=V_{\rm
mf}^{(s,v)}(q)+V_{\rm col}^{(s,v)}(q)$.  For a pion produced at $x$
with momentum $\kappa$ to propagate in the medium to the thermal
freeze-out point $x_f$ and be detected at the detecting point $x_d$
with momentum $k$, the probability amplitude according to the
path-integral method is \cite{Fey65,Kle05}
\begin{eqnarray}
K(\kappa x \to k x_d)
=\int {\cal D}q~ e^{ iS(\kappa x \to k x_d;q) },
\end{eqnarray}
where $\int \!\!{\cal D}q ...$ is the sum over all paths $ q $ from $x$
to $x_d$, and the action $S(\kappa x \to k x_d; q) $ is
\begin{eqnarray}
S(\kappa x \to k x_d;q)
=\int L(q,\dot {\bf q}) ~dt 
= - \int_x^{x_d, ({\rm path~}q)} p(q')\cdot dq'.
\end{eqnarray}
We can separate $S(\kappa x \to k x_d; q)$ into different contributions,
\begin{eqnarray}
\label{eq9}
S(\kappa x \to k x_d;q )
=-k\cdot (x_d - x) +\delta_{\rm mf} (\kappa x, k x_d; q)
+\delta_{\rm col}  (\kappa x, k x_d; q),
\end{eqnarray}
\vspace*{0.3cm}where
\vspace*{-1.2cm}
\begin{eqnarray}
\label{delmf}
\delta_{\rm mf} (\kappa x, k x_d; q)
=-\int_x^{x_d,({\rm path~}q)} [ p_{\rm mf}(q')-k ] \cdot dq'
=-\int_x^{x_f,({\rm path~}q)} [ p_{\rm mf}(q')-k ] \cdot dq',
\end{eqnarray}
\vspace*{-0.6cm}
\begin{eqnarray}
\label{eq11}
\delta_{\rm col} (\kappa x, k x_d;q)
=-\int_x^{x_d, ({\rm path~}q)} p_{\rm col}(q') \cdot dq'
=-\int_x^{x_f, ({\rm path~}q)} p_{\rm col}(q') \cdot dq',
\end{eqnarray}
\vspace*{-0.6cm}
\begin{eqnarray}
p_{\rm mf} ( q) 
=\biggl ( \gamma [ m_\pi+V_{\rm mf}^{(s)}( q)]+V_{\rm mf}^{0(v)}(q), 
~~\gamma [ m_\pi+V_{\rm mf}( q) ] \dot {\bf q} 
+{\bf V}_{\rm mf}^{(v)}(q)  \biggr ),  
\end{eqnarray}
\vspace*{-0.6cm}
\begin{eqnarray}
\label{eq13}
p_{\rm col}(q)
=\biggl ( \gamma  V_{\rm col}^{(s)} (q)+V_{\rm col}^{0(v)}(q), 
~~\gamma  V_{\rm col}^{(s)}(q) \dot {\bf q} 
+{\bf V}_{\rm col}^{(v)}(q) \biggr ).
\end{eqnarray}
Because of the additivity of the collision potentials in Eq.\
(\ref{col}), the phase shift for multiple collision $\delta_{\rm col}$
is a sum of the phase shifts for individual collisions, similar to the
case of the Glauber wave function in multiple scattering
\cite{Won03},
\begin{eqnarray}
\delta_{\rm col} (\kappa x, k x_d; q)
=\sum_i \delta_{{\rm col},i} (\kappa x, k x_d;q),
\end{eqnarray}
where $\delta_{{\rm col},i} (\kappa x, k x_d; q)$ is obtained from
Eqs.\ (\ref{eq11}) and (\ref{eq13}) with the potential $v_{\rm
col}(q-q_i)$ in place of the total collision potential $V_{\rm
col}(q)$.  The propagation amplitude is therefore
\begin{eqnarray}
K(\kappa x \to k x_d)
=\! \! \int \!\! {\cal D}q \exp \{  -ik\cdot(x_d -x)
+ i\delta_{\rm mf} (\kappa x, k x_d; q) 
+ i\delta_{\rm col}(\kappa x, k x_d; q) 
\}.
\end{eqnarray}
In this sum over all possible path $q$ in the above equation, we can
make the approximation that the dominant contribution to the path
integral comes from the trajectory along the classical path $q_c$ for
mean-field motion, as other trajectories give fluctuating
contributions that tend to cancel each other.  Then the amplitude is
approximately
\begin{eqnarray}
K(\kappa x \to k x_d)
\approx \exp \{  -ik\cdot(x_d -x)
+ i\delta_{\rm mf} (\kappa x, k x_d; q_c) 
+ i\delta_{\rm col}(\kappa x, k x_d; q_c) 
\}.
\end{eqnarray}
It is important to point out that the trajectory in the path integral
method is quite general and does not need to follow a straight line.
Our previous consideration for the propagation of an energetic pion
along a straight-line trajectory in Ref.\ \cite{Won03} is just a
special case of the path-integral method.

The wave amplitude for the pion to propagate from $x$ to $x_d$ is
\begin{equation}
\label{psi1}
\Psi (\kappa x \to k x_d)=A(\kappa x) e^{i\phi_0(x)} K(\kappa x \to 
k x_d)\,,
\end{equation}
where $A(\kappa x)$ is the production amplitude, $\phi_0(x)$ is a random 
production phase for chaotic source.

The probability amplitude for the production of two identical
pions $(\kappa_1, \kappa_2)$ at $(x_1, x_2)$ to be detected
subsequently as $k_1$ at $x_{d1}$ and $k_2$ at $x_{d2}$ is
\begin{eqnarray}
\label{psi2}
\frac{1}{\sqrt{2}}  \biggl \{ 
\Psi_1(\kappa_1 x_1 \to k_1 x_{d1}) \Psi_1(\kappa_2 x_2 \to k_2 
x_{d2}) + ( x_1 \leftrightarrow  x_2) \biggr \},
\end{eqnarray}
where $(x_1 \leftrightarrow x_2)$ represents the term symmetric to the
former by exchanging $x_1$ and $x_2$.  The probability $P(k_1,k_2)$
for the detection of two pions with momenta $(k_1, k_2)$ is the
absolute square of the sum of the two particle amplitudes
(Eq. (\ref{psi2})) from all $x_1$ and $x_2$ source points.  Because of
the random and fluctuating phase $\phi_0(x_i)$ for a chaotic source,
the absolute square of the sum of the amplitudes becomes the sum of
the absolute squares of the amplitudes \cite{Won94}.  One obtains
\begin{eqnarray}
P(k_1,k_2) = P(k_1)P(k_2)\, [1 + R(k_1 k_2)\,], 
\end{eqnarray}
where $P(k)$ is the probability of detecting one pion of momentum $k$, 
\begin{eqnarray}
\label{Pk1}
P(k)=\sum_x \big|A^2(\kappa x) K(\kappa x \to k x_d)\big|^2
=\sum_x e^{-2\,{\cal I}m\,\delta_{\rm col}(\kappa x, k x_{\rm f}; 
\,q_{\rm c})}A^2(\kappa x)\,,
\end{eqnarray}

\begin{eqnarray}
\label{PPR}
\hspace*{-15mm} P(k_1)P( k_2) R(k_1 k_2)
\!\!&=&\!\! \sum_{x_1 x_2} 
A(\kappa_1 x_1) A(\kappa_2 x_2) 
K(\kappa_1 x_1 \to k_1 x_{d1})K(\kappa_2 x_2 \to k_2 x_{d2})
\nonumber\\
& & \!\! \times
A(\kappa_1 x_2) A(\kappa_2 x_1) 
K^*(\kappa_1 x_2 \to k_1 x_{d1})K^*(\kappa_2 x_1 \to k_2 x_{d2}).
\end{eqnarray}
We can evaluate the product $K(\kappa_1 x_1 \to k_1 x_{d1})
K^*(\kappa_2 x_1 \to k_2 x_{d2})$.  It is equal to
\begin{eqnarray}
K(\kappa_1 x_1 \!\!\!&\to &\!\!    k_1 x_{d1})
K^*(\kappa_2 x_1 \to k_2 x_{d2})
= e^{-ik_1\cdot(x_{d1}-x_1)+ik_2\cdot(x_{d2}-x_2)} 
\nonumber\\
&&\!\!\!\!\times 
\int {\cal D}q \int {\cal D}q' 
\exp \{ 
 i\delta_{\rm mf}   (\kappa_1 x_1, k_1 x_{d1};q)
-i\delta_{\rm mf}^* (\kappa_2 x_1, k_2 x_{d2};q')
\nonumber\\
& &~~~~~~~~~~~~~~~~~~~~~~~~~~~+i\delta_{\rm col}  (\kappa_1 x_1, k_1 x_{d1};q)
-i\delta_{\rm col}^*(\kappa_2 x_1, k_2 x_{d2};q')
\}.
\end{eqnarray}
The real part of the phase difference $\delta_{\rm col} (\kappa_1 x_1,
k_1 x_{d1};q) -\delta_{\rm col}^*(\kappa_2 x_1, k_2 x_{d2};q')$ is
stationary when the trajectories $q$ and $q'$ inside the medium
coincide and is random and fluctuating when their trajectory in the
medium $q$ differs from $q'$.  Thus, we have
\begin{eqnarray}
&&\int {\cal D}q \int {\cal D}q' 
\exp \{ 
 i\delta_{\rm mf}   (\kappa_1 x_1, k_1 x_{d1};q)
-i\delta_{\rm mf}^* (\kappa_2 x_1, k_2 x_{d2};q')
\nonumber\\
& &~~~~~~~~~~~
+i\delta_{\rm col}  (\kappa_1 x_1, k_1 x_{d1};q)
-i\delta_{\rm col}^*(\kappa_2 x_1, k_2 x_{d2};q')
\}
\nonumber \\
&\sim&
\int {\cal D}q  
\exp \{ 
 i\delta_{\rm mf}   (\kappa_1 x_1, k_1 x_{d1};q)
-i\delta_{\rm mf}^* (\kappa_2 x_1, k_2 x_{d2};q)
\nonumber\\
& &~~~~~~~~~~
+i\delta_{\rm col}  (\kappa_1 x_1, k_1 x_{d1};q)
-i\delta_{\rm col}^*(\kappa_2 x_1, k_2 x_{d2};q)
\}.
\end{eqnarray}
Among all trajectories inside the medium, the dominant contribution to
the path integral comes from those along classical paths,
\begin{eqnarray}
& &\int {\cal D}q  
\exp \{ 
 i\delta_{\rm mf}   (\kappa_1 x_1, k_1 x_{d1};q)
-i\delta_{\rm mf}^* (\kappa_2 x_1, k_2 x_{d2};q)
\nonumber\\
& &~~~~~
+i\delta_{\rm col}  (\kappa_1 x_1, k_1 x_{d1};q)
-i\delta_{\rm col}^*(\kappa_2 x_1, k_2 x_{d2};q)
\}
\nonumber \\
&\sim &
\exp \{ 
 i\delta_{\rm mf}   (\kappa_1 x_1, k_1 x_{d1};q_c)
-i\delta_{\rm mf}^* (\kappa_2 x_1, k_2 x_{d2};q_c)
\nonumber\\
& &~~~~~
+i\delta_{\rm col}  (\kappa_1 x_1, k_1 x_{d1};q_c)
-i\delta_{\rm col}^*(\kappa_2 x_1, k_2 x_{d2};q_c)
\}.
\end{eqnarray}
As correlations occur for $k_1$ close to $k_2$ these classical paths
are close to each other, the real part of the collisional phase shifts
approximately cancel and only the additive imaginary part due to
absorption remain.  We then have
\begin{eqnarray}
i\delta_{\rm col} (\kappa_1 x_1, k_1 x_{d1};q_c) -i\delta_{\rm
  col}^*(\kappa_2 x_1, k_2 x_{d2};q_c) &\approx& -{\cal I}m
\delta_{\rm col}(\kappa_1 x_1, k_1 x_{d1}; q_c) -{\cal I}m
\delta_{\rm col}(\kappa_2 x_1, k_2 x_{d2}; q_c). 
\end{eqnarray}
The imaginary part of the multiple scattering phase shift ${\cal I}m
\delta_{\rm col} (k_1 x_1,k_1 x_{d1}; q_c)$ represents an absorption
along the trajectory and can be simplified to be
\cite{Won03,Won04,Won05,Zha04c}
\begin{eqnarray}
\label{pk123}
2 {\cal I}m ~\delta_{\rm col}(\kappa_1 x_1, k_1 x_{d1}; q_c)
=\int_{x_1}^{x_f,({\rm path~} q_c)} n_{\rm med}(q')~\sigma_{\rm abs} ~dq',
\end{eqnarray}
where $n_{\rm med}(q')$ is the medium density at $(q')$, $\sigma_{\rm
abs}$ is the pion absorption cross section, and $x_f$ is the
freeze-out point. As correlations occur when $k_1$ is close to $k_2$
and the absorptions for $k_1$ and $k_2$ from $x_1$ along the
trajectory inside the medium are nearly the same, ${\cal I}m
\delta_{col} (\kappa_1 x_1, k_1 x_{d1}; q_c)$ is approximately equal
to ${\cal I}m \delta_{col} (\kappa_2 x_1, k_2 x_{d1}; q_c)$.  We can
abbreviate these quantities respectively as ${\cal I}m \delta_{col}
(k_1,x_1)$ and ${\cal I}m \delta_{col} (k_2,x_1)$ with the other
variables implicitly understood.  

For the mean-field part of the phase shift, we have
\begin{eqnarray}
\delta_{\rm mf}   (\kappa_1 x_1, k_1 x_{d1};q_c)
-\delta_{\rm mf}^* (\kappa_2 x_1, k_2 x_{d2};q_c)
&=&\int_{x_1}^{x_f,({\rm path~}q_c)} \{ [p_{1\, {\rm mf}} (q)-k_1] 
- [p_{2\, {\rm mf}}^*(q)-k_2] \} \cdot dq
\nonumber\\
&\equiv&
\phi_{\rm mf}(k_1 k_2,x_1; q_c)
\end{eqnarray}
                                                                                
Using $\int d^4x \rho(x)$ to substitute $\sum_x$ in Eqs. (\ref{Pk1}) and
(\ref{PPR}) for a continuous pion-emitting source with a density $\rho(x)$, 
we have
\begin{eqnarray}
\label{pk1}
P(k) = \int\!d^4x \rho(x)\,e^{-2 {\cal I}m\,{\delta}_{\rm col}(k,x)
} A^2(\kappa x)\,,
\end{eqnarray}
\begin{eqnarray}
\label{pk12}
P(k_1) P(k_2) R(k_1,k_2)
= \int d^4x_1 d^4x_2 \,\rho(x_1) \rho(x_2)\, e^{-2\,{\cal I}{m}\,
{\delta}_{\rm col}(k_1,x_1)} e^{-2\,{\cal I}{m}\,{\delta}_{\rm col}
(k_2,x_2)} \big|\Phi(x_1 x_2; k_1 k_2)\big|^2 ,
\end{eqnarray}
\begin{eqnarray}
R(k_1,k_2)= \biggl |
\int d^4x e^{i(k_1-k_2)\cdot x +i\phi_{\rm mf}(k_1k_2,x; q_c)
-{\cal I}m \delta_{\rm col}(k_1,x) - {\cal I}m \delta_{\rm col}(k_2,x)}
\rho_{\rm eff}(k_1k_2,x) \biggr |^2 ,
\end{eqnarray}
where
\begin{eqnarray}
\label{PHI}
\Phi(x_1 x_2; k_1 k_2) = \frac{1}{\sqrt{2}} \bigg\{{\bar A}(x_1,\kappa_1
k_1) {\bar A}(x_2,\kappa_2 k_2) e^{ik_1\cdot x_1 +ik_2\cdot x_2} +
(x_1 \leftrightarrow x_2) \bigg\}\,,
\end{eqnarray}
\begin{eqnarray}
{\bar A}(x,\kappa k)= A(\kappa x)\,e^{i\delta_{\rm mf}
(\kappa x \to k x_{\rm f}; \,q_{\rm c})},
\end{eqnarray}
\begin{eqnarray}
\rho_{\rm eff}(k_1 k_1,x)
=\frac {\sqrt{f(\kappa_1 x)f(\kappa_2 x)}} {P(k_1)P(k_2)},
\end{eqnarray}
and $f(\kappa x)=\rho(x)A^2(\kappa x)$ is the pion-emitting function
of the chaotic source at $x$ with momentum $\kappa$ that evolves
asymptotically to $k$ \cite{Won03}.  We obtain the simple result that
the multiple scattering of the interfering pair leads essentially to a
factor containing the imaginary part of the phase shift, which is
related to the absorption of the particles through the medium
\cite{Won03,Won04,Won05,Zha04c}.  Thus, even though the source expands
to a greater spatial dimension during the late stage from chemical
freeze-out to thermal freeze-out, the initial chaotic source at
chemical freeze-out remains an important element in determining the
pair correlation function.

We shall calculate the absorption factor in subsections IIIC and
discuss its effect on HBT results in subsection IIID.  The collective
flow (mean-field interaction) `distorts' the initial momentum $\kappa$
into the final detected momentum $k$.  The phase associated with the
collective expansion of the source, ${\phi}_{\rm mf}(k_1 k_2,x;q_c)$,
is is very small \cite{Won03}.  Our calculations indicate that its
effect on HBT results can be neglected.

\subsection{Pion-emitting sources}

In our model calculations for the two-pion HBT correlation functions,
the identical pions (for example $\pi^+$) are assumed to be emitted
thermally from the space-time hypersurface at $T^{\rm ch}$ for CFO
production and from the hypersurface at $T^{\rm th}$ for TFO
production.  For PCE case, the identical pions includes primary pions
produced at the chemical freeze-out (CFO production) as well as
secondary pions arising from decay products of excited-state particles
during the chemical freeze-out to thermal freeze-out.  The
four-dimension density of the pion source for PCE production can be
written as
\begin{equation}
\rho(x)=n_{\pi}(x)\delta(\tau-\tau^{ch})+\sum_{j\ne \pi}D_{j\rightarrow \pi}
n_j(x)\,,
\end{equation}
where $\tau^{ch}$ is the chemical freeze-out time in local frame and
$D_{j\rightarrow \pi}$ is the decay rate $\Gamma_j$ times 
its corresponding fraction ${\tilde d}_{j\rightarrow \pi}$.  For
example, $D_{\Delta \rightarrow \pi}= \Gamma_{\Delta}\times
\frac{1}{3}$, $D_{\rho \rightarrow \pi}=\Gamma_{\rho} \times
\frac{2}{3}$, $D_{\omega \rightarrow \pi}=\Gamma_{\omega} \times
0.88$, and $D_{\pi^0\pi^0 \rightarrow \pi^+\pi^-}=v_r n_{\pi}
\sigma(\pi^0\pi^0 \rightarrow \pi^+\pi^-) \times 1$, where $v_r$ is
the relative velocity of the two colliding pions.  In our
calculations, the values of $\Gamma_{\Delta}$, $\Gamma_{\rho}$, and
$\Gamma_{\omega}$ are taken as 120 MeV, 145 MeV, and 8 MeV
\cite{Hag02}.  The cross section $\sigma(\pi^0\pi^0
\rightarrow \pi^+\pi^-)$ is equal to the absorption cross section,
$\sigma_{\rm abs}(\pi^+\pi^- \rightarrow \pi^0\pi^0)$, and will be
discussed in the next subsection.

\begin{figure}
\vspace*{-1.0cm}
\includegraphics*[angle=0,width=8.0cm,height=8.0cm]{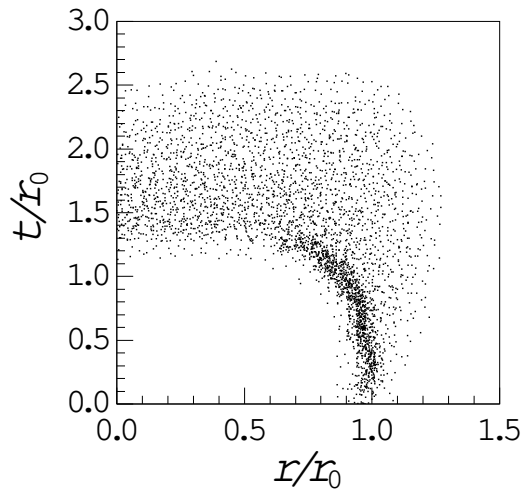}
\vspace*{-0.5cm}
\caption{\label{fig:fig2}Space-time distribution of the pion-emitting 
source for PCE production at AGS energies. }
\end{figure} 

\begin{figure}
\vspace*{-1.0cm}
\includegraphics*[angle=0,width=8.0cm,height=8.0cm]{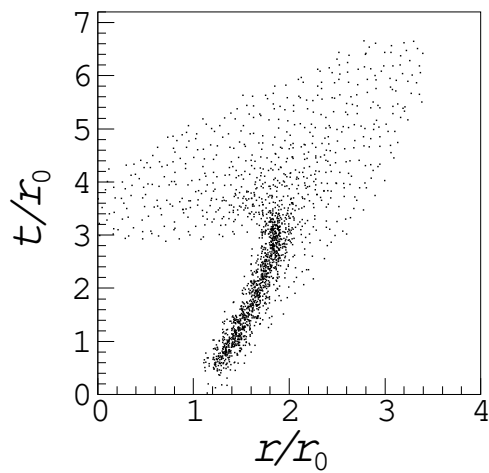}
\vspace*{-0.5cm}
\caption{\label{fig:fig3}Space-time distribution of the pion-emitting 
source for PCE production at RHIC energy. }
\end{figure} 

Fig. 2 shows the space-time distribution of the pion-emitting source
in our calculations for the PCE production at AGS energies.  Fig. 3
shows the space-time distribution of the pion-emitting source for the
PCE particle production at RHIC energy.  The produced primary pion
source points are distributed near the center of the source and the
decay product secondary pion source points are distributed in the
outer region of the source.  One can see that the space-time geometry
of the sources produced at AGS and RHIC energies are different.

\subsection{Cross sections of pion absorption}

For the PCE case, the pions propagating in the source during the
chemical freeze-out to the thermal freeze-out will subject to multiple
scattering with the medium particles in the source.  Based on
Glauber multiple scattering theory \cite{Gla59}, the absorption factor
owing to multiple scattering in Eqs. (\ref{pk1}) and (\ref{pk12}) can
be written as \cite{Won03,Won04,Won05,Zha04c}
\begin{eqnarray}
\label{absf}
e^{-2\,{\cal I}{m}\,{\delta}_{\rm col}(x,k)}=\exp\Bigg[\!-
\!\int_{x}^{x_{\rm f}}\! {\sum_i}'\sigma_{\rm abs}(\pi i)\, 
n_i(q_{\rm c})\, d\ell(q_{\rm c})\Bigg],
\end{eqnarray}
where $\sum'_i$ means the summation for all particles in the source except 
for the test pion, $\sigma_{\rm abs}(\pi i)$ is the absorption cross section 
of the pion with the medium particle $i$, and $n_i(q_{\rm c})$ is the 
density of the medium particle $i$ along the classical propagation path 
$q_{\rm c}$.  

For the source produced at RHIC energy, pions are the dominant
component of produced particles.  The other produced particles
considered in our model are $K$, $\eta$, $\rho$, and $\omega$.
Because the threshold of the reaction $\pi \pi \rightarrow K {\bar K}$
is greater than $\langle \sqrt{s_{\pi \pi}}\,\rangle$ in the source
and the probability of three-pion collision for $\pi \pi \pi
\rightarrow \omega$ is very small, the dominant contributions for the
multiple scattering absorption are from reactions $\pi^+ \pi \to \rho$
and $\pi^+\pi^- \to \pi^0 \pi^0$.  In our model, the cross sections
for the reactions $\pi^+ \pi \to \rho$ and $\pi^+\pi^- \to \pi^0$ are
calculated by \cite{Mar76,Ynd02}
\begin{equation}
\label{sigma1}
\sigma_{\rm abs}(\pi^+ \pi \to \rho)=\frac{4 \pi}{p_{\rm cm}^2}\sin^2\delta_1
\end{equation}
and
\begin{equation}
\label{sigma2}
\sigma_{\rm abs}(\pi^+\pi^-\rightarrow\pi^0\pi^0)={8\over 9}{\pi\over
{p_{\rm cm}^2}} \sin^2(\delta_0-\delta_2)\,,
\end{equation}
where $p_{\rm cm}=\sqrt{s_{\pi\pi}-4m_{\pi}^2}/2$ and the phase shifts
$\delta_0$, $\delta_1$, and $\delta_2$ are given by Ref. \cite{Ynd02}.

For the source produced at AGS energies, the dominant contribution for the
multiple scattering absorption is from the process $\pi^+ N \rightarrow 
\Delta$.  We calculate the absorption cross section in our model by
\begin{equation}
\sigma_{\rm abs}(\pi^+ N\rightarrow\Delta)=\frac{2}{3} \frac{\sigma_0\,
(\Gamma_{\Delta}/2)^2}{(\sqrt{s_{\pi N}}-m_{\Delta})^2+
(\Gamma_{\Delta}/2)^2}\,,
\end{equation}
where $m_{\Delta}=1232$ MeV and $\sigma_0=200$ mb.

\subsection{Calculation of $P(k)$ and $P(k_1,k_2)$}

After knowing the hydrodynamical solution, we have the space-time 
distributions of temperature, chemical potential, and velocity of 
the evolving source.  Then, we can calculate $P(k)$ and $P(k_1,k_2)$ 
for constructing HBT correlation function based on Eqs. (\ref{pk1}) and
(\ref{pk12}) in the following steps: 

Step 1: Select space-time coordinate of the first test pion, $x_1$, in 
the source randomly with the probability of the space-time source density 
$\rho(x)$ (at $T^{\rm ch}<T<T^{\rm th}$, see Figs. 2 and 3).

Step 2: Generate the momentum of the pion in local frame, $\kappa_1'
=(E_1',\bkappa_1')$, according to the Bose-Einstein distributions at the 
point $x_1$, which is a function of the temperature and chemical potential 
at $x_1$.

Step 3: Boost $\kappa_1'$ by the Lorentz transform with the source fluid 
velocity at $x_1$.  Then, we get the momentum of the pion in the center of 
mass frame of the source, $\kappa_1=(E_2,\bkappa_1)$.

Step 4: Determine the new space coordinate of the pion propagating along 
classical path at next time by the production of the pion's velocity 
$\bbeta_1=\bkappa_1/E_1$ and the step of time $\Delta t$.  If the 
temperature at the new point is higher than $T^{\rm ch}$ we re-select 
the pion from step 1 to step 3. 

Step 5: Boost $\kappa_1'$ by the Lorentz transform with the source 
velocity at the new point to get the momenta of the pion at the point 
and calculate $w_{\rm ms1}=\sum_j \sigma_{\rm abs}(\pi j)n_j 
\sqrt{(\bbeta_1 \Delta t)^2}$ at this new point.

Step 6: Repeat steps 4 and 5 until the point $x_f$ at the temperature
$T^{\rm th}$ to get the final (detected) momentum of the pion, $k_1$, 
and accumulate $w_{\rm ms1}$ to get the values of $e^{-2\,{\cal I}{m}\,
{\delta}_{\rm col}(x_1,k_1)}$ based on Eq. (\ref{absf}).

Step 7: Repeat step 1 to step 6 to get the corresponding quantities for 
the second test pion.  

Step 8: According to Eqs. (\ref{pk1}) and (\ref{pk12}) calculate the 
weights
\begin{equation}
W_i=e^{-2\,{\cal I}{m}\,{\delta}_{\rm col}(x_i,k_i)} 
[E_i'/E_i({\bf k}_i)]\,,
~~~~~~i=1,2
\end{equation}
and
\begin{eqnarray}
W_{12}&=&e^{-2\,{\cal I}{m}\,{\delta}_{\rm col}(x_1,k_1)} 
e^{-2\,{\cal I}{m}\,{\delta}_{\rm col}(x_2,k_2)} \,\frac{1}{2}\Big |
\sqrt{E'_1/E_1}\sqrt{E'_2/E_2}
\nonumber\\
&\times&e^{ik_1\cdot x_1 +ik_2\cdot x_2} +\sqrt{E'_{12}/E_1}
\sqrt{E'_{21}/E_2} \,e^{ik_1\cdot x_2 +ik_2\cdot x_2} \Big |^2
\end{eqnarray}
for $P(k_i)$ $(i=1,2)$ and $P(k_1,k_2)$, where $E'_{ij}\,(i,j=1,2)$ 
is the energy of the $i$th pion in the local frame at $x_j$, which is 
obtained from $E_i$ by a reverse Lorentz transformation with the source 
velocity at $x_j$.  

Step 9: Repeat step 1 to step 8 and accumulate $W_i~(i=1,2)$ and $W_{12}$ 
for each $({\bf k}_1, {\bf k}_2)$ bin, respectively.  Then we get the 
one-pion and two-pion momentum distributions $P(k)$ and $P(k_1,k_2)$ 
for constructing HBT correlation functions.    

\subsection{HBT results}

The two-particle HBT correlation function $C(k_1,k_2)$ is defined as the
 ratio of the two-pion momentum distribution $P(k_1,k_2)$ to the the product
 of the single-pion momentum distribution $P(k_1) P(k_2)$,
 \begin{equation}
 \label{c2}
 C(k_1,k_2) = {{P(k_1,k_2)}\over {P(k_1)P(k_2)}}\,.
 \end{equation}
Using the relative momentum of the two selected pions, $q=|{\bf k_1} -
{\bf k_2}|$, as variable, we construct the two-pion correlation function
$C(q)$ from $P(k_1,k_2)$ and $P(k_1)P(k_2)$ by picking the identical pion
pairs emitted from the source and summing over ${\bf k_1}$ and ${\bf k_2}$
for each $q$ bin \cite{Zha93,Zha04c}.  Fig. 4 (a) and (b) show the two-pion
correlation functions for sources produced at AGS and RHIC energies,
respectively.  In Fig. 4, the symbols circle and up-triangle symbols are 
the results for the CFO and PCE particle productions.  The square symbols 
present the results for the TFO particle production, for this case the HBT 
calculations of quantum path-integral reduce to a usual HBT calculation 
for a certain source configuration.  The curves in Fig. 4 are obtained by 
fitting the model calculation data of the two-pion correlation functions 
with the parametrized formula
\begin{equation}
\label{C} C(q)=1+\lambda e^{-q^2R^2}\,.
\end{equation}

\begin{figure}
\includegraphics*[angle=0,scale=0.45]{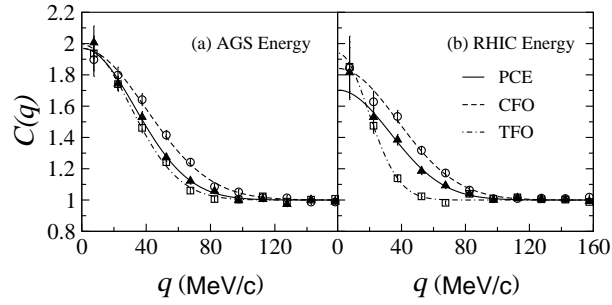}
\caption{\label{fig:fig4}The two-pion correlation functions for the sources
produced at AGS and RHIC energies.}
\end{figure}

Table 1 gives the fitted HBT radii $R$ and chaotic parameter $\lambda$
in this parametrized formula.  In our model calculations, the initial
source radius, $r_0$, is taken as 6 fm both for the sources produced
at AGS and RHIC energies.  We can see that the HBT radii for the CFO
and PCE productions are obviously smaller than those for the TFO
particle production, especially at RHIC energy.  This arises because
that the thermal freeze-out is the latest states of the expanding
sources, which have the largest space geometry of the sources, and at
RHIC energy the expanding velocity is higher.  In order to investigate
the effects of the excited-state particle decays and multiple
scattering absorption on the HBT results in the PCE case, we also
examine the HBT correlation functions without the multiple-scattering
absorptions by letting the absorption factor equal to 1.  The
corresponding HBT results are listed on right column of Table 1.  One
can see that the HBT radii for the PCE without multiple-scattering
absorptions case are larger than those for the CFO case because of the
excited-state particle decays.  The multiple scattering of the
identical pions with the medium particles in the source leads to an
absorption factor to the source density for calculating the two-pion
HBT correlation functions.  It constrains the pions produced near the
center of the source and leads to a larger space configuration of the
pion-emitting source.

By comparing the HBT radii for the PCE case with and without
absorption for the source produced at AGS energies, Table 1 indicates
that the effect of the large absorption by delta formation increases
the HBT radius, as the source of pions are located farther out in the
expanding system.  On the other hand, the HBT radius for the PCE case
is close to that for the PCE case without absorption, for collisions
at RHIC energies.  This indicates that for RHIC collisions, the
effects of absorption is small because the cross sections for
$\pi^+\pi^- \to \pi^0\pi^0$ and $\pi\pi \to \rho$ are small.  Our
results show that at RHIC energy the values of $\lambda$ for the CFO
and PCE cases are smaller than unit even for the chaotic sources we
consider in the paper.  Investigating the reason of the $\lambda$
results will be of interest in our future work.

\begin{table*}
\caption{\label{tab:table1}The HBT fitted results (the unit of source
radii $R$ is fm).}
\begin{ruledtabular}
\begin{tabular}{ccccc}
&PCE&CFO&TFO&PCE non-abs.\\
\hline
AGS&$R=4.20\pm0.15$&$R=3.51\pm0.11$&$R=4.64\pm0.16$&$R=3.79\pm0.13$\\
&$\lambda=0.97\pm0.05$&$\lambda=0.97\pm0.05$&$\lambda=0.99\pm0.06$&
$\lambda=0.98\pm0.05$\\
RHIC&$R=4.22\pm0.23$&$R=3.74\pm0.15$&$R=7.28\pm0.42$&$R=4.44\pm0.23$\\
&$\lambda=0.70\pm0.06$&$\lambda=0.84\pm0.06$&$\lambda=0.94\pm0.12$&
$\lambda=0.79\pm0.07$\\
\end{tabular}
\end{ruledtabular}
\end{table*}

\section{Summary and discussion}

Quantum transport of the pair of interfering pion pair has an
important influence in the properties of HBT interferometry in
heavy-ion collisions.  Staring with a chaotic source at chemical
freeze-out, we examine the effects of this quantum transport using the
path integral method, in which the evolution of the bulk matter is
described by relativistic hydrodynamics while the paths of the two
interfering pions are described by test particles following the
hydrodynamical fluid positions and velocity fields.  By following the
trajectories of the interfering pion pair after chemical freeze-out up
to thermal freeze-out, we find that the quantum transport leads to a
cancellation of the real part of the phase shifts but the imaginary
part leads to an absorption of the propagating pions.  The transport
gives rise to an a HBT radius closer to the radius at chemical
freeze-out.  We examine in addition the effects of particle decay from
which secondary pions are produced after chemical freeze-out.  As the
system continue to expand after chemical freeze-out, the spatial
dimension of the secondary pions are greater than those of the
chemical freeze-out source.  Secondary pions will lead to a greater
HBT radius.  The combined effects of of multiple scattering and
excited-state particle decays lead to HBT radii for the PCE case
larger than those for the CFO case, but smaller than those of the TFO
case.

The nature of the particle-emitting sources produced in high energy
heavy ion collisions at RHIC energy is different from that produced in
the collisions at AGS energies.  At AGS energies, the sources may be
described by mixed hadronic gas with finite baryon densities.  The
main stable particles in the sources are nucleons and produced pions,
and the main excited-state particle may be considered to be $\Delta
(1232)$.  Whereas at RHIC energies, the sources undergo a phase
transition from the quark-gluon plasma (QGP) phase to the hadronic gas
phase, and with an approximate zero baryon density.  The main stable
particles are produced pions and kaons, and there are many
excited-state particles, such as $\rho$ and $\omega$ in the sources.
At AGS energies this effect increases the HBT radius for the PCE case
to be substantially larger than the HBT radius for the CFO case.
Whereas at RHIC energies, the HBT radius for the PCE case increases
only slightly from the CFO value and is about the same as the PCE case
without absorption, because the absorption cross sections are small.

In our calculations we consider only the spherical evolving sources
for simplicity.  In high-energy heavy-ion collisions, the evolution of
the system is different in the longitudinal direction (beam direction)
and the transverse direction.  This effect is important for 
sources produced at RHIC energies.  Therefore, it will be of great
interest to examine the HBT interferometry with quantum transport of
the interfering pion pair for more reasonable evolving source models
and investigate the HBT radii in different directions (``out",
``side", and ``long" directions) in the future.

\begin{acknowledgments}
This research was supported by the National Natural Science Foundation of
China under Contract Nos. 10575024 and 10775024 and in part by the Division
of Nuclear Physics, US DOE, under Contract No. DE-AC05-00OR22725 managed by
UT-Battle, LC.
\end{acknowledgments}

\end{document}